\documentstyle[aps,graphicx]{revtex}

\begin{document}
\preprint{\vbox{\hbox{UCD-99-6} }}
\draft
\title {Direct Signals of Low Scale Gravity at $e^+ e^-$ Colliders}
\author{Kingman Cheung}
\address{ 
Department of Physics, University of California, Davis, 
CA 95616 USA}
\author{Wai-Yee Keung}
\address{ 
Department of Physics, University of Illinois at Chicago, 
Chicago IL 60607-7059}
\date{\today}
\maketitle

\begin{abstract}
Gravity can become strong at the TeV scale in the theory of extra dimensions.
An effective Lagrangian can be used to describe the gravitational interactions
below a cut-off scale.  In this work, we study the associated production of
the gravitons with a $Z$ boson or a photon at $e^+ e^-$ colliders of energies
of LEPII to the Next Linear Colliders (NLC) ($\sqrt{s}=0.25-1.5$ TeV) and 
calculate the sensitivity to the new interactions.  We also obtain the limit 
on the cut-off scale using the present data from LEPII.
\end{abstract}

\section{Introduction}

Recent advances in string theories suggest that  a special
11-dimension theory (dubbed as M theory) \cite{mtheory} may be the theory of 
everything.  Impacts of M theory on our present world can be studied with
compactification of the 11 dimensions down to our $3+1$ dimensions.  The 
path of compactification is, however, not unique.  One popular path is to first
compactify the 11 dimensions down to 5 dimensions \cite{5d}.  In this
5-dimensional world, the standard model particles live on a brane ($3+1$ dim)
while there are other fields, like gravity and super Yang-Mill fields, 
live in the bulk.  A novel mechanism to break the supersymmetry (SUSY) 
is offered
in this picture, in which a hidden sector lives on another brane in this 
5-dimensional world.  This brane is entirely separated from the standard
model (SM) brane.
While SUSY is broken in this hidden brane by any means, the SUSY breaking
is communicated to the SM brane via the interactions of the fields in
the bulk.  Large number of studies in this area have been proposed 
\cite{msusy}.
Apart from the above, radical ideas like TeV scale string theories 
were also proposed \cite{tevstring}.

Inspired by string theories a simple but probably workable solution to
the gauge hierarchy was recently 
proposed by Arkani-Hamed, Dimopoulos and Dvali (ADD)
\cite{theory}.  They assumed the space is $4+n$ dimensional, with the
SM particles living on a brane.  While the electromagnetic, strong,
and weak forces are confined to this brane, gravity can also propagate 
in the extra dimensions.  To solve the gauge hierarchy problem they
proposed the ``new'' Planck scale $M_S$ is of the order of 
TeV in this picture with
the extra dimensions of a very large size $R$.
The usual Planck scale 
$M_G=1/\sqrt{G_N} \sim 1.22 \times 10^{19}$ GeV is related to this effective
Planck scale $M_S$ by using Gauss's law:
\begin{equation}
R^n \, M_S^{n+2} \sim M_G^2 \;.
\end{equation}
For $n=1$ it gives a large value for $R$, which is already ruled out 
by gravitational
experiments.  On the other hand, $n=2$ gives $R \alt 1$ mm, which is 
in the margin beyond the reach of present gravitational experiments.
  
The graviton including its excitations in the extra dimensions 
can couple to the SM particles on the brane with a strength of
$1/M_S$ at short distances 
(instead of $1/M_G$), and thus the gravitation interaction
becomes comparable in strength to weak interaction at TeV scale.  Hence, it
can give rise to a number of phenomenological activities testable at existing
and future colliders \cite{wells,shrock,han,mira,collider}.  
So far, studies show that there
are two categories of signals: direct and indirect.  The indirect signal
refers to exchanges of gravitons in the intermediate states, while direct
refers to production or associated production of gravitons in the final 
state.  
Since gravitons interact weakly with detectors, they will escape detection
and give rise to missing energies.  Thus, the logical signal to search for 
would be the associated production of gravitons with other SM particles.  
At $e^+ e^-$ colliders, the best signals would be the associated production
of graviton with a $Z$ boson, a photon, or a fermion pair.
The production of graviton and photon at LEPII 
has been studied in Ref. \cite{mira}.
In this work, we shall study the associated production of graviton with
a $Z$ boson or a photon at $e^+ e^-$ colliders of energies from LEPII
to about 1.5 TeV.

The detection of photon will be direct while for the 
$Z$ boson its decay products, a lepton pair or a quark pair, need to be 
detected.  The branching ratio of the $Z$ boson into visible products is 
about 80\% (the decay of $Z$ into neutrinos 
will not be useful because the final state would then be all missing
energies.)  Therefore, the signature will be a lepton pair or a quark pair
of the $Z$ mass with a large missing energy (or a photon with missing energy 
for $\gamma G$ production.)
In contrary to the background, the 
recoil mass spectrum will not show a particular resonance because the 
mass spectrum of the graviton excitations is almost a continuous one.
Recent analyses were performed by L3, ALEPH, and DELPHI \cite{aleph} 
at LEPII in search for $ZH$ production with the Higgs boson decaying into 
invisible particles.  This signature is similar to our signal of interest.  
Thus, we shall also use these data to constrain the new cut-off scale $M_S$.
In order to have a handle on the feasibility of our signal, we choose 
$ZH$ production as a bench-mark for comparison with the cross section of 
$ZG$.  
The major background comes from $e^+ e^- \to Z\;(\gamma)\; \nu_i \bar\nu_i 
\;(i=e, \mu,\tau)$, of which $\nu_\mu$ and $\nu_\tau$ come mainly from
$ZZ\;(\gamma  Z)$ pair production while $\nu_e$ also has contributions from 
$W$-exchange diagrams.  

The organization of the paper is as follows.  In the next section, we shall
give the details of the calculations. In Sec. III, we shall give numerical
results and derive possible limits that can be reached by the NLC.  
In Sec. IV, we use the data from LEPII on the search for $ZH$ production with
the Higgs boson decaying invisibly and the data on the production of 
a single photon 
with missing energy to constrain the cut-off scale.
We shall then conclude in Sec. V.

\section{Calculations}

We concentrate on the spin-2 component of the Kaluza-Klein (KK) 
states, which are the excited modes of  graviton in the extra dimensions.
The spin-1 and spin-0 components are less interesting 
phenomenologically. We follow the convention  in Ref.\cite{han}.
There are four contributing Feynman diagrams for the process $e^+ e^- \to
Z G$: one 4-point vertex diagram and the other three diagrams are obtained
by attaching the graviton to each leg of $e^+$, $e^-$, and $Z$.  
The amplitudes for $e^-(p_1) e^+(p_2) \to Z(k_1) G(k_2)$  are given by:
\begin{eqnarray}
{\cal M}_1 &=& \frac{g \kappa}{2 \cos\theta_W}\, \frac{1}{s - m_Z^2} \,
\bar v(p_2) \gamma_\beta (g_v- g_a\gamma^5) u(p_1) \; 
\left( \eta^{\alpha\beta} - \frac{ (k_1+k_2)^\alpha (k_1+k_2)^\beta }{m_Z^2}
\right )
\nonumber \\
&\times & \biggr [
 - 2 k_1\cdot k_2 \, \epsilon^\nu (k_1) \epsilon_{\alpha\nu}(k_2) 
+ 2 \epsilon^\nu (k_1) \epsilon_{\mu\nu}(k_2) k_1^\mu {k_1}_\alpha \nonumber \\
&&- 2 \epsilon_\alpha (k_1) \epsilon_{\mu\nu}(k_2) k_1^\mu k_1^\nu
+ 2 k_2\cdot \epsilon(k_1) \epsilon_{\alpha\mu}(k_2) k_1^\mu 
            \biggr ] \;, \\
{\cal M}_2 &=& \frac{g \kappa}{2 \cos\theta_W}\, \frac{1}{(k_2-p_2)^2} 
\bar v(p_2) \gamma^\mu (\overlay{/}{k}_2 - \overlay{/}{p}_2 ) 
\overlay{/}{\epsilon}(k_1) (g_v - g_a \gamma^5) u(p_1) \; p_2^\nu \; 
\epsilon_{\mu\nu}(k_2) \;, \\
{\cal M}_3 &=& -\frac{g \kappa}{2 \cos\theta_W}\, \frac{1}{(p_1-k_2)^2} 
\bar v(p_2) \overlay{/}{\epsilon}(k_1) (g_v - g_a \gamma^5) \,
(\overlay{/}{p}_1 - \overlay{/}{k}_2 ) 
\gamma^\mu u(p_1) \; p_1^\nu \; \epsilon_{\mu\nu}(k_2) 
\; \\
{\cal M}_4 &=& \frac{g \kappa}{2 \cos\theta_W} \; \bar v(p_2) \gamma^\nu 
(g_v - g_a \gamma^5) \,u(p_1) \; \epsilon^\mu(W) \, \epsilon_{\mu\nu}(G) \;. 
\end{eqnarray}
where $\kappa =\sqrt{16\pi G_N}$, $g_v$ and $g_a$ are the vector and 
axial-vector coupling of $Z$ to electron.

We have used the REDUCE program 
to evaluate the square of the sum of  amplitudes.
The spins of the $Z$ boson and the graviton are summed using these formulas:
\begin{eqnarray}
\sum_s \epsilon^s_\mu(k_1) \, \epsilon^{s*}_\nu(k_1) &=&
 -\eta_{\mu\nu} + \frac{ {k_1}_\mu  {k_1}_\nu}{ m_Z^2} \;, \nonumber \\
\sum_s \epsilon^s_{\mu\nu}(k_2) \, \epsilon^{s*}_{\rho\sigma}(k_2) &=&
 \hbox{$1\over2$} B_{\mu\nu,\rho\sigma}(k_2) \; \;,
\end{eqnarray}
where $B_{\mu\nu,\rho\sigma}(k_2)$ is given by\cite{han}
$$
B_{\mu\nu,\rho\sigma}(k_2)=
P_{\mu\rho,\nu\sigma}(k_2)+P_{\mu\sigma,\rho\nu}(k_2)
-\hbox{$2\over3$}P_{\mu\nu,\rho\sigma}(k_2) \ ,
$$
\begin{equation}
P_{\mu\nu,\rho\sigma}(k_2)=(\eta_{\mu\nu}-k_{2\mu} k_{2\nu}/m^2)
                           (\eta_{\rho\sigma}-k_{2\rho} k_{2\sigma}/m^2)
\ .
\end{equation}
Here $m$ denotes the mass of the KK state.
The spin-averaged amplitude squared for the process 
$e^+ e^- \to ZG$ is given by
\begin{eqnarray}
\overline{\sum} |{\cal M}|^2 &=& \frac{g^2 \kappa^2 (g_v^2 + g_a^2)}
{48 \cos^2 \theta_W u^2 t^2
(s-m_Z^2)^2} \; \Biggr \{
8 m_Z^6 t u [ 3 m^2 ( m^2 -t-u) + 4 t u ] \nonumber \\
&& + 2 m_Z^4 tu [ 27 m^6 - 42 m^4 (t+u) + 15 m^2 ( t^2 + u^2) + 80 m^2 t u
  - 28 ( t^2  u + t  u^2 ) ] \nonumber \\
&& + m_Z^2 [ 3m^8 ( -t^2 -u^2 + 12 t u)
       +6 m^6 (t^3 - 12( t^2 u + t u^2) +u^3 )
       +3 m^4 (-t^4 +14 t^3 u +62 t^2 u^2 + 14 t u^3 - u^4 ) \nonumber \\
&& \qquad       +6 m^2 ( -t^4 u -23(t^3 u^2 + t^2 u^3) -t u^4)
       + 36 (t^4 u^2 + t^2 u^4) + 52 t^3 u^3  ] \nonumber \\
&& + 3ut ( -m^2 +t+u) [ -m^4 +m^2(t+u) -4tu][ 2m^4 -2 m^2 (t+u) +t^2 +u^2 ]
 \;\Biggr \} \;.
\end{eqnarray}
To obtain the total cross section we have to sum over all discrete KK states 
with $m_k = 2\pi k/R$ for all $m_k$ below $\sqrt{s} - m_Z$.
Since the mass spacing of these KK states is much smaller than any physical
scales in the problem, it is convenient to convert the discrete sum on $k$
to an integral over $m^2$ as follows:
\begin{equation}
\sum_k \Rightarrow \int dm^2 \; \frac{R^n m^{n-2} }{ (4\pi)^{n/2} \Gamma(n/2)}
\;.
\end{equation}
The size $R$ of the extra dimension, the scale $M_S$ and $G_N$ are
related by
\footnote{The definition of the cut-off scale $M_S$ (we follow Ref. \cite{han})
is different from the cut-off scale $M$ of Ref. \cite{mira}.  $M_S$ is 
related to the $M$ by $M_S^4 = 4M^4$ for $n=2$.
}
\begin{equation}
G_NR^nM_S^{n+2}=(4\pi)^{n\over2}\Gamma(n/2) \;.
\end{equation}

For the similar process $e^+ e^- \to \gamma G$ we can reproduce the 
expression given in Ref. \cite{mira}
(for unpolarized case):
\begin{eqnarray}
\frac{d\sigma}{d \cos\theta} &=& \frac{\pi \alpha G_N}{ 4 \left( 1- 
\frac{m^2}{s} \right )} \; \biggr[ (1+\cos^2\theta) \left( 1 +
(\frac{m^2}{s})^4 \right ) \nonumber \\
&&+ \left( \frac{ 1-3\cos^2\theta +4\cos^4\theta}{1-\cos^2\theta} \right )
\, \frac{m^2}{s} \, \left( 1+ (\frac{m^2}{s})^2 \right ) +
6 \cos^2\theta (\frac{m^2}{s})^2 \biggr ] \;.
\end{eqnarray}

\section{Cross sections and Distributions}

\subsection{$e^+ e^- \to ZG$}

\begin{figure}[th]
\leavevmode
\begin{center}
\includegraphics[height=4.5in]{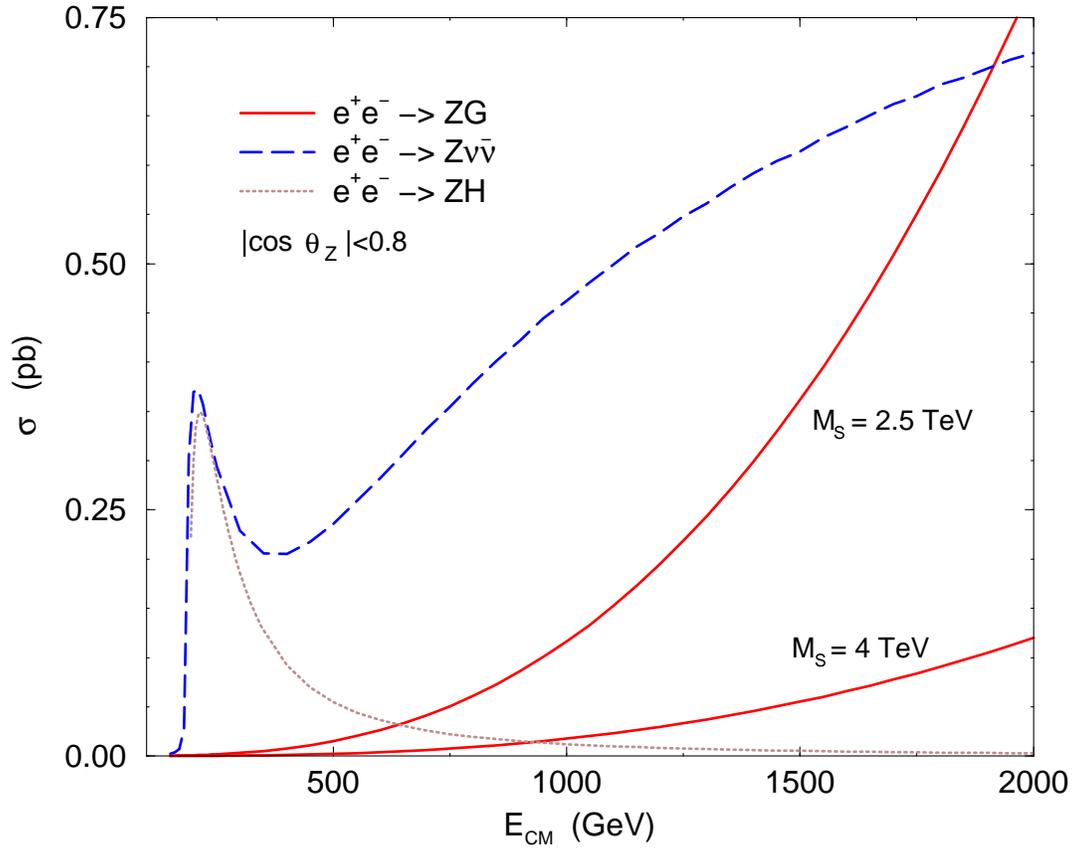}
\end{center}
\caption{The total cross sections for $e^+ e^- \to Z \nu_i \bar\nu_i (i=e,\mu,
\tau)$, $e^+ e^- \to ZG$, and the $e^+ e^- \to ZH$ with $|\cos\theta_Z|<0.8$.
We used $n=2$ and $M_S=2.5$ and 4 TeV as shown.
 }
\label{fig1}
\end{figure}

We start with the result for the associated production of graviton with
the $Z$.
The total cross sections for the signal ($e^+ e^- \to ZG$) and background
($e^+ e^- \to Z \nu \bar \nu$) versus $\sqrt{s}$ for $n=2$ and $M_S=2.5,4$
TeV with an angular cut $|\cos \theta_Z|<0.8$
are shown in Fig. \ref{fig1}, where we
also show the bench-mark process $e^+ e^- \to ZH$.  
The angular cut, though is not necessary, can help reducing the background.
The background contains three flavors of neutrinos.  
The $Z\nu_\mu \bar \nu_\mu$ and 
$Z\nu_\tau \bar \nu_\tau$ production mainly comes from $ZZ$ production and
decreases with $\sqrt{s}$, while 
$Z \nu_e \bar \nu_e$ can also come from the 
$t-$channel $W$-exchange diagrams and so increases with $\sqrt{s}$.
The signal cross section increases with $\sqrt{s}$ because more KK levels
contribute to the cross section.
Due to phase space suppression the signal cross section is rather small at 
low $\sqrt{s}$.  Only
until $\sqrt{s}$ reaches at least 0.5 TeV does the signal become significant
relative to the background.

\begin{figure}[th]
\leavevmode
\begin{center}
\includegraphics[height=4in]{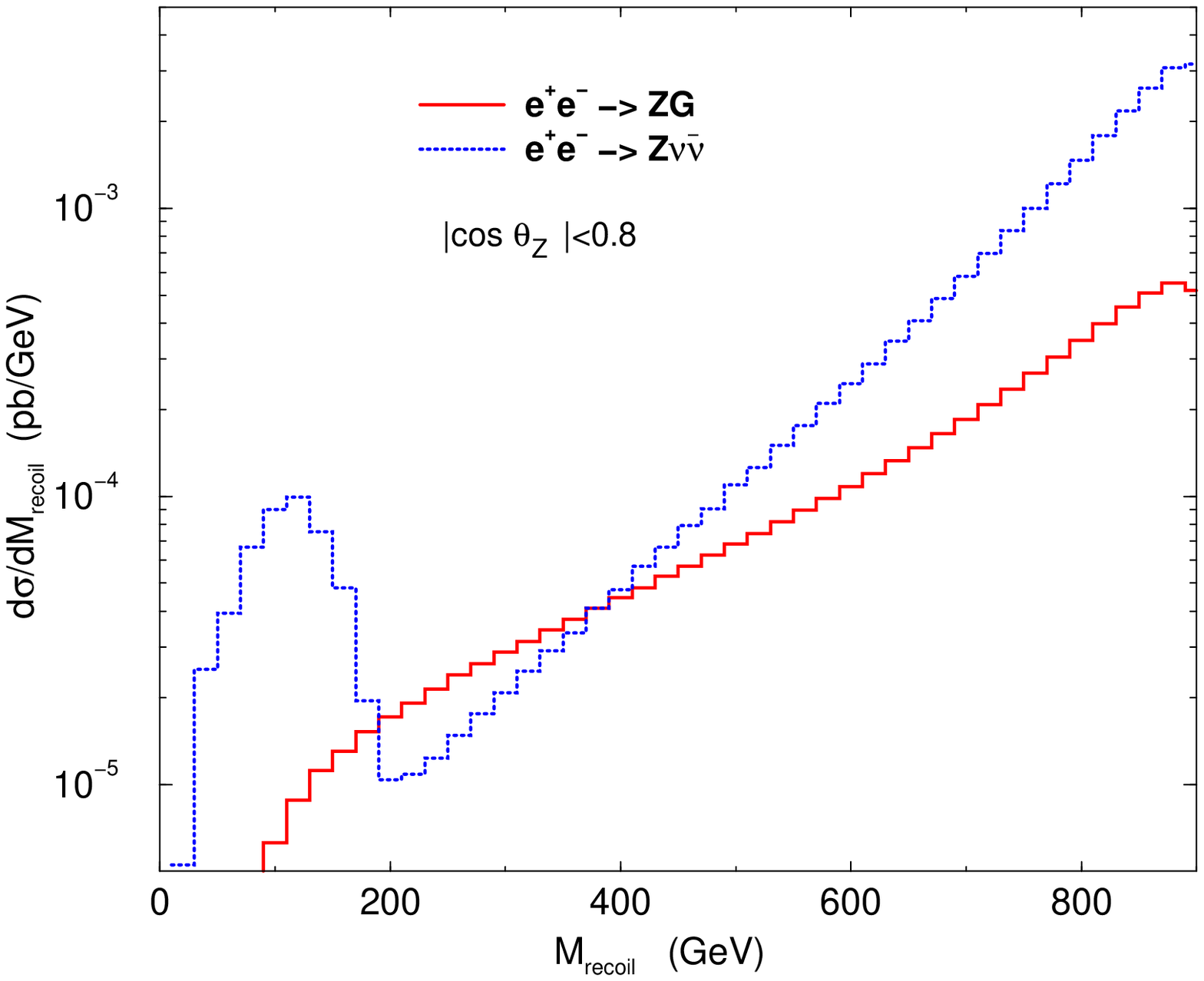}
\includegraphics[height=4in]{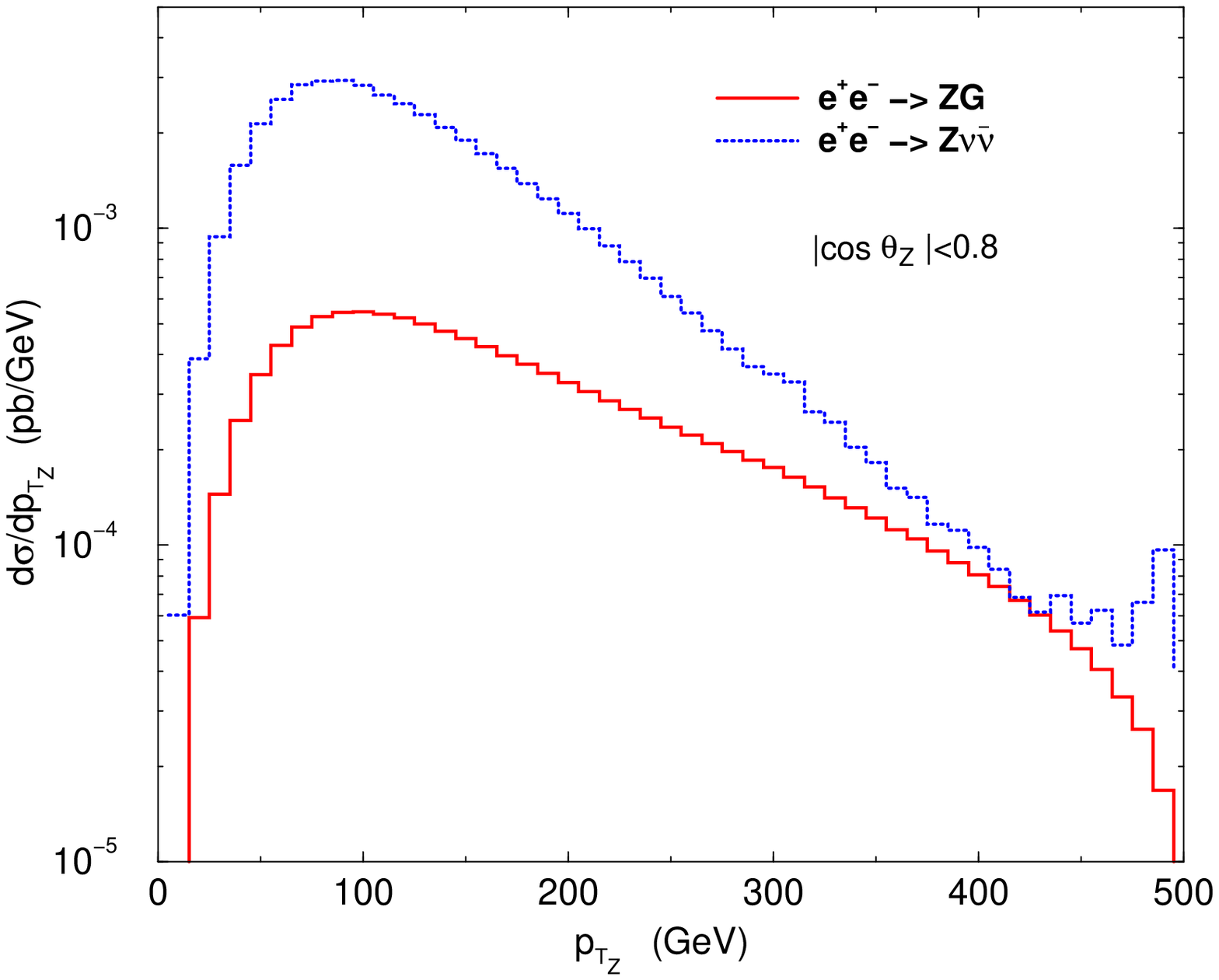}
\end{center}
\caption{(a) The differential distribution $d\sigma /d M_{\rm recoil}$ and
(b) $d\sigma /d P_{T_Z}$ 
for $e^+ e^- \to Z \nu_i \bar\nu_i$ $(i=e,\mu, \tau)$ 
and $e^+ e^- \to ZG$
for $n=2$ and $M_S=2.5$ TeV.  
We imposed a cut of $|\cos\theta_Z|<0.8$.
 }
\label{fig2}
\end{figure}

We look more closely at a particular $\sqrt{s}$ and examine kinematic
distributions and see if we can find some ways to improve the 
signal-to-background ratio.  We choose $\sqrt{s}=1$ TeV, $n=2$, and
$M_S=2.5$ TeV.  
The $ZZ$ production is rather back-to-back while the $ZG$ is
less  back-to-back, and that is why we imposed a cut on the angle of the 
$Z$ boson, which will not hurt the signal too much.  
In reality, the $Z$ boson actually
decays visibly into either a pair of quarks or leptons with a branching 
ratio of $0.8$.  Experimental reconstruction of the $Z$ boson is excellent
and so in this study we only impose a smearing on the $Z$ momentum to
approximate the decay.  
We have used an energy resolution of $\delta E/E = 0.2/\sqrt{E}$ 
for the $Z$ boson, which gives approximately a 15--20 GeV spread of
the reconstructed $Z$ mass.  In Fig. \ref{fig2} we use $M_S=2.5$ TeV.  In
Ref. \cite{wells}, the effective theory with a cut-off at $M_S$ remains
valid up to a few times of $M_S$ as long as unitarity is concerned. 
Perhaps, the gravitation interaction already becomes strong at or below the
scale $M_S$. 

\begin{table}[t]
\caption{\label{table1}
The signal $S$, background $B$, signal-to-background ratio 
$S/B$, and the significance $S \sqrt{{\cal L}}/\sqrt{B}$ for  $e^+ e^- \to
ZG$ at $e^+ e^-$ colliders of various $\sqrt{s}$ with a luminosity ${\cal L}$
of 50 fb$^{-1}$ for $n=2$ and $M_S=2.5$ TeV. 
Cuts of $|\cos\theta_Z|<0.8$ and $M_{\rm recoil} >200$ GeV are imposed.}
\medskip
\begin{tabular}{|ccccc|}
$\sqrt{s}$ (TeV) & $S$ (fb) & $B$ (fb)& $S/B$ & $S\sqrt{\cal L}/\sqrt{B}$ \\
\hline
0.5   & 13.1   & 179 & 0.073 &  6.9  \\
0.75  & 48.9   & 334 & 0.15  &  18.9 \\
1.0   & 115    & 452 & 0.25  &  38.1 \\
1.25  & 217    & 543 & 0.40  &  65.8 \\
1.5   & 360    & 613 & 0.59  &  103 \\
\end{tabular}
\end{table}

Another useful variable is the recoil mass, $M_{\rm recoil}$, which is 
defined as
\begin{equation}
M_{\rm recoil} = \biggr[ s - 2 \sqrt{s} E_Z + m_Z^2({\rm recons.})
  \biggr]^{1/2} \;.
\end{equation}
In Fig. \ref{fig2}(a) we compare the recoil mass spectrum of the signal with
that of the background.  Obviously, a part of the background comes from
$ZZ$ production and, therefore, the visible $Z$ boson recoiled against
the another $Z$ boson.  It explains a peak around 90 GeV.  The spectrum for
the signal, on the other hand, does not show any peak structure.  This
is a characteristics of the continuous mass spectrum of the KK levels.  
A cut of 
\begin{equation}
M_{\rm recoil} > 200  \;\;{\rm GeV}
\end{equation}
can remove the $ZZ$ background.  However, the rest of the recoil mass
spectra for the signal and background look very much alike.  
Another useful distribution is the transverse momentum of the visible $Z$
boson.  The background falls more rapidly than the signal, which means we
can impose a cut on the transverse momentum to increase the 
signal-to-background ratio: see Fig. \ref{fig2}(b).  From Fig. \ref{fig2}(b)
a cut of about 150 GeV may help  but, however, it also cuts away about half 
of the signal.  Therefore, we do not impose any cuts on $p_{T_Z}$.

\begin{table}[t]
\caption{\label{table2}
Table showing the limit of $M_S$ that can be reached at $e^+ e^-$ 
colliders of various $\sqrt{s}$.
The signal $S$, $S/B$, and the significance $S \sqrt{{\cal L}}/\sqrt{B}$ 
are also shown.  A luminosity ${\cal L}$ of 50 fb$^{-1}$ is assumed.
Part (a) requires both the significance larger than 5 and $S/B >0.1$ for 
discovery while part (b) only requires significance larger than 5.
Cuts of $|\cos\theta_Z|<0.8$ and $M_{\rm recoil} >200$ GeV
are imposed.}
\medskip
\begin{tabular}{|ccccc|}
&&&&\\
\multicolumn{5}{|c|}{(a) With $S/B>0.1$ and $S\sqrt{\cal L}/\sqrt{B} >5$ }\\
&&&&\\
$\sqrt{s}$ (TeV) & $M_S$ limit (TeV) &  $S$ (fb) & $S/B$  & 
$S\sqrt{\cal L}/ \sqrt{B}$  \\
\hline
0.5   & 2.3   & 17.9 & 0.1  &  9.5  \\
0.75  & 2.8   & 33.4 & 0.1  &  12.9 \\
1.0   & 3.2   & 45.2 & 0.1  &  15.0 \\
1.25  & 3.5   & 54.3 & 0.1  &  16.5 \\
1.5   & 3.9   & 61.3 & 0.1  &  17.5 \\
\hline
\hline
&&&&\\
\multicolumn{5}{|c|}{(b) With $S\sqrt{\cal L}/\sqrt{B} >5$} \\
&&&&\\
\hline
0.5   & 2.7   & 9.5  & 0.053  &  5 \\
0.75  & 3.5   & 12.9 & 0.039  &  5 \\
1.0   & 4.2   & 15.0 & 0.033  &  5 \\
1.25  & 4.8   & 16.5 & 0.030  &  5 \\
1.5   & 5.3   & 17.5 & 0.029  &  5 \\
\end{tabular}
\end{table}

We show the cross sections for the signal $S$ and the background $B$, the 
signal-to-background ratio $S/B$, and the significance 
$S\sqrt{\cal L}/\sqrt{B}$ at $e^+ e^-$ colliders of energies from 0.5 to 1.5
TeV, with the choice of $n=2$ and $M_S=2.5$ TeV, in Table \ref{table1}.  
The nominal yearly luminosity
at these next-linear colliders is of the order of 50 fb$^{-1}$.  With such a
luminosity a decent amount of signal with a large significance is achievable
for $n=2$ and $M_S=2.5$ TeV.  Since the cross section for the signal scales
as $1/M_S^4$ for $n=2$ and $1/M_S^6$ for $n=4$, and even higher power of 
$1/M_S$ for larger $n$, so the signal cross section drops rapidly with $n$ or
$M_S$.  For $n>2$ the signal cross section in Table \ref{table1},
being further down by some orders of magnitudes, becomes phenomenologically
uninteresting.  Thus, we concentrate on the case $n=2$.  
In Table \ref{table2}, we show the limit on the cut-off scale $M_S$ that
can be obtained by requiring both the $S/B>0.1$ and significance larger than
5.  The limit ranges from about 2.3 TeV to about 4 TeV.  
If we only require the significance of the signal larger than 5 to set the 
limit, the limit on $M_S$ is slightly better, ranging from 2.7 TeV to about 
5.3 TeV.  However, in this case  the signal-to-background ratio 
becomes too small.

There are two  other important backgrounds, which are 
$e^+ e^- \to W^+ W^-$ and $W^\pm\ell^\mp \nu$ production when the $W$ boson(s) 
decays
leptonically.  If the invariant mass of the lepton pair falls within the
$Z$ mass region, 
such events will look like a $Z$ boson with missing energies.  
Fortunately, these two backgrounds are reducible if (i) we use only
the hadronic decay of the $Z$ boson in our signal, or (ii) restrict the 
reconstructed $Z$ mass to a narrower range.  
The semi-leptonic decay mode of $WW \to q \bar q' \ell \nu$ will be 
relevant as a true background to $ZG$ only if the measurement of the 
$q\bar q'$ invariant mass falls within the $Z$ mass region and the lepton
$\ell$ is missing down the beam pipe.  
These two conditions  reduce the $W^+ W^-, W^\pm \ell^\mp \nu$ backgrounds
to a negligible level.  Thus, the remaining concern is the leptonic decay
mode of $W^+ W^-$ and $W^\pm \ell^\mp \nu$ if we include the leptonic mode of
$Z$ in our signal.  We performed a calculation of $e^+ e^- \to 
W^\pm \ell^\mp \nu$ with $W^\pm \to \ell^\pm \nu \;(\ell =e,\mu)$, and
impose the requirement that the invariant mass of $\ell^+ \ell^-$ falls
within the $Z$ mass region (80--120 GeV) and impose also the other cuts:
$|\cos \theta_Z|<0.8$, $M_{\rm recoil}>200$ GeV.
We found that these $W^+ W^-, W^\pm \ell^\mp \nu$ backgrounds are only 
about 5\% of the $Z\nu\bar \nu$ background at $\sqrt{s}=0.5$ TeV and
continuously decreases to 0.5\% at $\sqrt{s}=1$ TeV and finally less than
0.1\% at $\sqrt{s}=1.5$ TeV.  We can, therefore, safely ignore these 
backgrounds.

\subsection{$e^+ e^- \to \gamma G$}

\begin{figure}[th]
\leavevmode
\begin{center}
\includegraphics[height=4.5in]{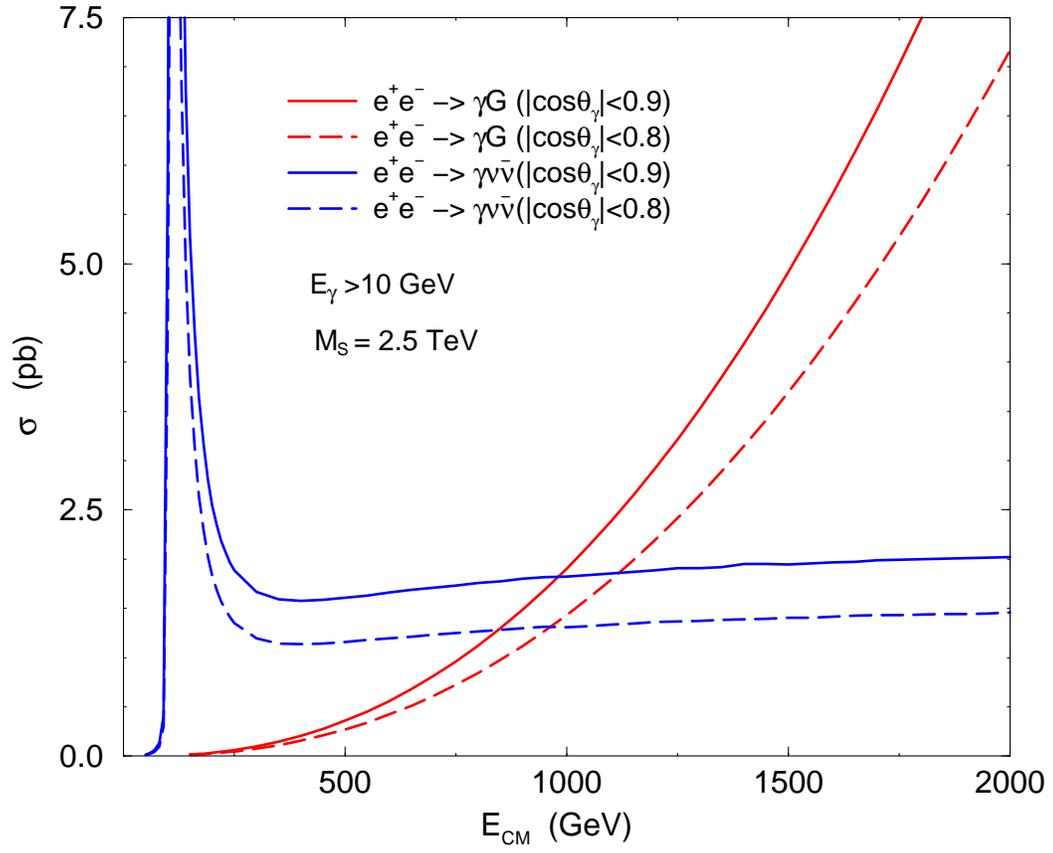}
\end{center}
\caption{The total cross sections for $e^+ e^- \to \gamma \nu_i \bar\nu_i 
(i=e,\mu, \tau)$ and  $e^+ e^- \to \gamma G$ with cuts as shown. 
We used $n=2$ and $M_S=2.5$ TeV.
 }
\label{fig3}
\end{figure}

Here we repeat the study on the process $e^+ e^- \to \gamma G$ with the 
background $e^+ e^- \to \gamma \nu \bar \nu$.  The cross sections for the 
signal and background versus $\sqrt{s}$ for $n=2$ and $M_S=2.5$ TeV 
with $|\cos\theta_\gamma|<0.9 \;(0.8)$ and $E_\gamma>10$ GeV are shown in
Fig. ~\ref{fig3}.  The signature of the event is a single photon with
missing energy in the final state.  For this signature the signal cross
section easily becomes larger than the background cross section when
$\sqrt{s}$ reaches  about 1 TeV.  Again, we show the recoil mass and
photon transverse momentum spectrum for the signal and the background.
They show similar characteristics as the $ZG$ channel.
In table \ref{table3}, we show the signal and background cross sections for
$n=2$ and $M_S=2.5$ TeV at energies from 189 GeV to 1.5 TeV.  At LEPII 189 GeV
energy, the signal is only about 3.2\% of the background and has a 
significance of only 0.65.  Therefore, any effect of the 
new gravity interactions with $M_S$ of the order of 
2.5 TeV is unnoticeable.  Finally, in table \ref{table4}
we show the limit of $M_S$ that can be obtained for $\sqrt{s}=189$ GeV to
1.5 TeV, by requiring both $S/B>0.1$ and the significance larger than 5. 
The limit that can be obtained at LEPII is only about 1.5 TeV while at 
$\sqrt{s}=1.5$ TeV it can reach up to about 5.6 TeV. 

So overall, the limit on the cut-off scale $M_S$ ranges from about 1.5
TeV to 5.6 TeV for $\sqrt{s}$ from LEPII energy to 1.5 TeV at
$e^+ e^-$ colliders using the associated production of graviton with a
$Z$ boson or a photon. Such a small increase in the limit is due to the 
scaling of the cross section with high powers of $1/M_S$.

\begin{table}[t]
\caption{\label{table3}
The signal $S$, background $B$, signal-to-background ratio 
$S/B$, and the significance $S \sqrt{{\cal L}}/\sqrt{B}$ for  $e^+ e^- \to
\gamma G$ at $e^+ e^-$ colliders of various $\sqrt{s}$ and a
luminosity ${\cal L}$ of 50 fb$^{-1}$ (0.5 fb$^{-1}$ at LEPII) for $n=2$ and
$M_S=2.5$ TeV. 
Cuts of $|\cos\theta_\gamma|<0.9$, $E_\gamma > 10$ GeV,
and $M_{\rm recoil} >200$ GeV (120 GeV at
LEPII and $\sqrt{s}=0.25$ TeV) are imposed.}
\medskip
\begin{tabular}{|ccccc|}
$\sqrt{s}$ (TeV) & $S$ (fb) & $B$ (fb)& $S/B$ & $S\sqrt{\cal L}/\sqrt{B}$ \\
\hline
0.189 & 26.2   & 815 & 0.032  &  0.65  \\
0.25  & 59.5   & 972 & 0.061  &  13.5  \\
0.5   & 352    & 1406 & 0.25  &  66.5  \\
0.75  & 955    & 1652 & 0.58  &  166 \\
1.0   & 1898   & 1791 & 1.06  &  317 \\
1.25  & 3211   & 1878 & 1.71  &  524 \\
1.5   & 4912   & 1939 & 2.53  &  789 \\
\end{tabular}
\end{table}

\begin{table}[t]
\caption{\label{table4}
Table showing the limit of $M_S$ that can be reached at 
$e^+ e^-$ colliders of various $\sqrt{s}$ 
for the process $e^+ e^- \to \gamma G$.
The signal $S$, $S/B$, and the significance $S \sqrt{{\cal L}}/\sqrt{B}$ 
are also shown.  A luminosity ${\cal L}$ of 50 fb$^{-1}$ (0.5 fb$^{-1}$ 
at LEPII) is assumed.
It requires both the significance larger than 5 and $S/B >0.1$ for 
discovery. Cuts of $|\cos\theta_\gamma|<0.9$, $E_\gamma > 10$ GeV,
and $M_{\rm recoil} >200$ GeV
(120 GeV at LEPII and $\sqrt{s}=0.25$ TeV) are imposed.}
\medskip
\begin{tabular}{|ccccc|}
&&&&\\
\multicolumn{5}{|c|}{(a) With $S/B>0.1$ and $S\sqrt{\cal L}/\sqrt{B} >5$ }\\
&&&&\\
$\sqrt{s}$ (TeV) & $M_S$ limit (TeV) &  $S$ (fb) & $S/B$  & 
$S\sqrt{\cal L}/ \sqrt{B}$  \\
\hline
0.189 & 1.5   & 202  & 0.25  &  5  \\
0.25  & 2.2   & 97.3 & 0.1  &  22.0  \\
0.5   & 3.1   & 141 & 0.1  &  26.6  \\
0.75  & 3.9   & 165 & 0.1  &  28.7 \\
1.0   & 4.5   & 179 & 0.1  &  29.9 \\
1.25  & 5.1   & 188 & 0.1  &  30.6 \\
1.5   & 5.6   & 194 & 0.1  &  31.1 
\end{tabular}
\end{table}

\section{Comparison with Existing data}

\subsection{$e^+ e^- \to ZG$}

As in the last section, we put in the cross section for the process 
$e^+ e^- \to ZH$ as a comparison to the signal of $e^+ e^- \to ZG$.  These
two processes share the same signature when the Higgs boson decays 
invisibly \cite{higgs}. Recently,  L3, ALEPH, and DELPHI \cite{aleph}
have searched for the invisibly decaying Higgs in association with a $Z$ 
boson.  We shall
use their  data to constrain the cut-off scale $M_S$.

The best limit on $M_H$ with the Higgs boson decaying invisibly was obtained by
ALEPH \cite{aleph} in their recent analysis at LEPII with
$\sqrt{s}=189$ GeV.  The 95\%CL lower limit on $M_H$ is 87.5 GeV (only
preliminary) assuming the Higgs boson is produced with the SM strength
in the $ZH$ production and the Higgs boson decays 100\% invisibly.
We calculate the corresponding 95\%CL upper limit on the cross section 
$\sigma(e^+ e^- \to ZH)= 0.55$ pb in the leading-order.  
With the following two approximations we can apply this limit directly to 
$e^+ e^- \to ZG$:  (i)
since the reported selection efficiencies are rather stable for a 
fairly wide range of $M_H$ (we refer to the similar analysis done at 
$\sqrt{s}=181-184$ GeV, the details at $\sqrt{s}=189$ GeV are not available), 
we can treat the limit of $\sigma(e^+ e^- \to ZH)=0.55$ pb as roughly
constant over a wide range of $M_H$. (ii) 
The signal of $e^+ e^- \to ZG$ has the same signature as 
$e^+ e^- \to ZH$ and so they should have similar efficiencies.  
We can then apply this cross section limit to the signal of $ZG$ and we
obtain the 95\%CL limit on $M_S$:
\begin{equation}
M_S \agt 515\;{\rm GeV}\;.
\end{equation}
This estimate is only a rough estimate but should be sufficient for our
purpose.  Even if there were a large difference between the selection
efficiencies for $ZG$ and $ZH$ (which is not likely), the change in 
$M_S$ would still be small, because the cross section scales as $1/M_S^4$ for
$n=2$.  

\subsection{$e^+ e^- \to \gamma G$}

The LEP Collaborations \cite{lep}
have been searching for single-photon events with
missing energies.  This is an interesting signature for a number of new
physics, including supersymmetry, which are (i) $e^+ e^- \to XY, X \to
Y\gamma$, (ii) $e^+ e^- \to \tilde{G} \tilde{G} \gamma$, and (iii) $e^+ e^-
\to \tilde{G} \tilde{\chi}^0_2, \tilde{\chi}^0_2 \to \tilde{\chi}^0_1 \gamma$.
Their limits on the single-photon cross section are rather model-dependent
because the detection efficiencies depend on model parameters.  
Since in our case it is very difficult to fully simulate the experimental 
conditions, we simply use their 95\%CL upper limits on production cross
sections.  It means we assume that the detection efficiencies for our graviton
signal are within the ranges of these experiments.  
We show the limits on the cut-off scale $M_S$ in table \ref{table6}. 
In the table, we can see that the limits reported by the LEP Collaborations
have rather wide ranges, simply because of the wide range of the detection
efficiencies.  It justifies our assumption that the efficiencies of
our graviton signal are easily within the ranges of these LEP experiments.
The limits we obtain are from about 1.2 to 2.2 TeV
(it is consistent with the value of $M=1.2$ TeV obtained in Ref. \cite{mira},
please see the footnote.)

\begin{table}[t]
\caption{\label{table6}
The range of the 95\%CL upper limit on the single photon production
cross section at $\sqrt{s}=183$ GeV (the data by DELPHI is preliminary at 
$\sqrt{s}=189$ GeV)  given by the LEP Collaborations and the corresponding 
limits of the cut-off scale obtained from these measurements.}
\medskip
\begin{tabular}{cc}
95\%CL limit of single-photon cross section  &  95\%CL limit on $M_S$ \\
\hline
 & \\
L3 & \\
$E\gamma>5$ GeV, $|\cos\theta_\gamma|<0.97$, $\sigma_{95} \sim 0.1-0.5$ pb &
$\sim 1.5 - 2.2$ TeV \\
\hline
& \\
ALEPH & \\
$p_{T_\gamma}>0.0375 \sqrt{s}$, $|\cos\theta_\gamma|<0.95$, 
$\sigma_{95} \sim 0.1-0.6$ pb &
$\sim 1.2 - 1.9$ TeV \\
\hline
&\\
OPAL & \\
$x_T=p_{T_\gamma}/E_{\rm beam}>0.05$, $|\cos\theta_\gamma|<0.966$, 
$\sigma_{95} \sim 0.075-0.8$ pb &
$ \sim 1.2 - 2.2$ TeV \\
\hline
&\\
DELPHI & \\
$x_\gamma=E_\gamma/E_{\rm beam}>0.06$, $45^\circ<\theta_\gamma<135^\circ$,
$\sigma_{95} \sim 0.3-0.4$ pb & 
$\sim 1.25 - 1.35 $ TeV
\end{tabular}
\end{table}

\section{Conclusions}

Excess signals of missing energy events in the process  
$e^+e^- \to Z(\gamma) +\overlay{/}{E}$ can provide a useful test for
the low scale gravity with extra space dimension compactified at the size
of mm. This suggests a TeV $e^+e^-$ collider can investigate the
direct graviton production and study the early unification.
Using the available data at LEPII a limit of $M_S \agt 515$ GeV is obtained 
in the $ZG$ channel
while $M_S \agt 1.2-2.2$ TeV is obtained in the $\gamma G$ channel.
For the future $e^+ e^-$ linear colliders of energies $0.25-1.5$ TeV and
a luminosity of 50 fb$^{-1}$ $M_S$ can be probed up to about  5.6 TeV.
Although some other non-accelerator physics, e.g., cooling of supernova 
\cite{theory},
may give a much stronger constraint on the cut-off scale $M_S$,
the collider signatures, including those considered in this paper, 
would provide independent tests for the new gravity interactions.

\section*{\bf Acknowledgments}
This research was supported in part by the U.S.~Department of Energy under
Grants Nos. DE-FG02-84ER40173 and DE-FG03-91ER40674 and
by the Davis Institute for High Energy Physics. 
K.C. wants to thank the National Center for Theoretical Sciences in Taiwan
where part of this work was done.

\begin{figure}[th]
\leavevmode
\begin{center}
\includegraphics[height=4in]{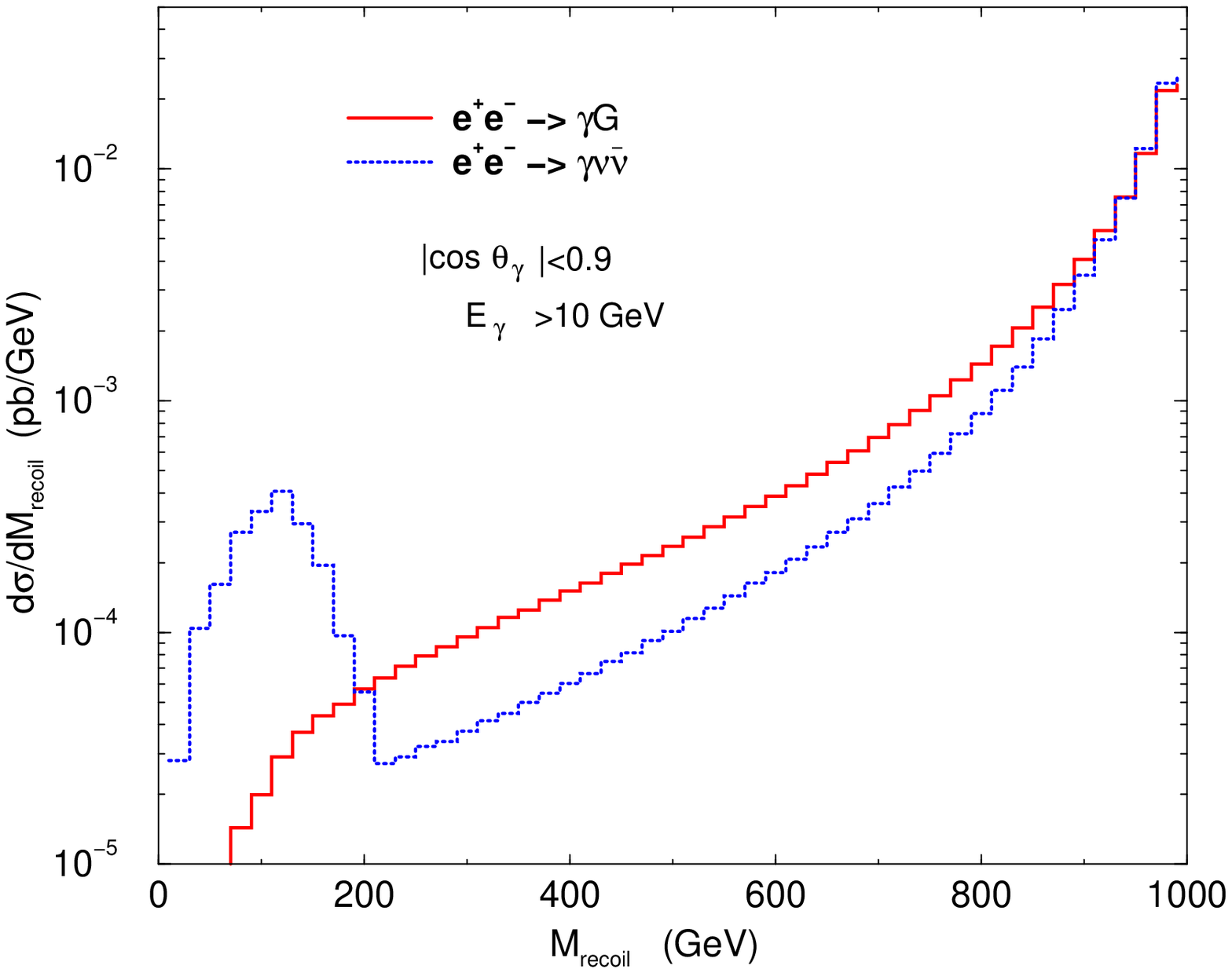}
\includegraphics[height=4in]{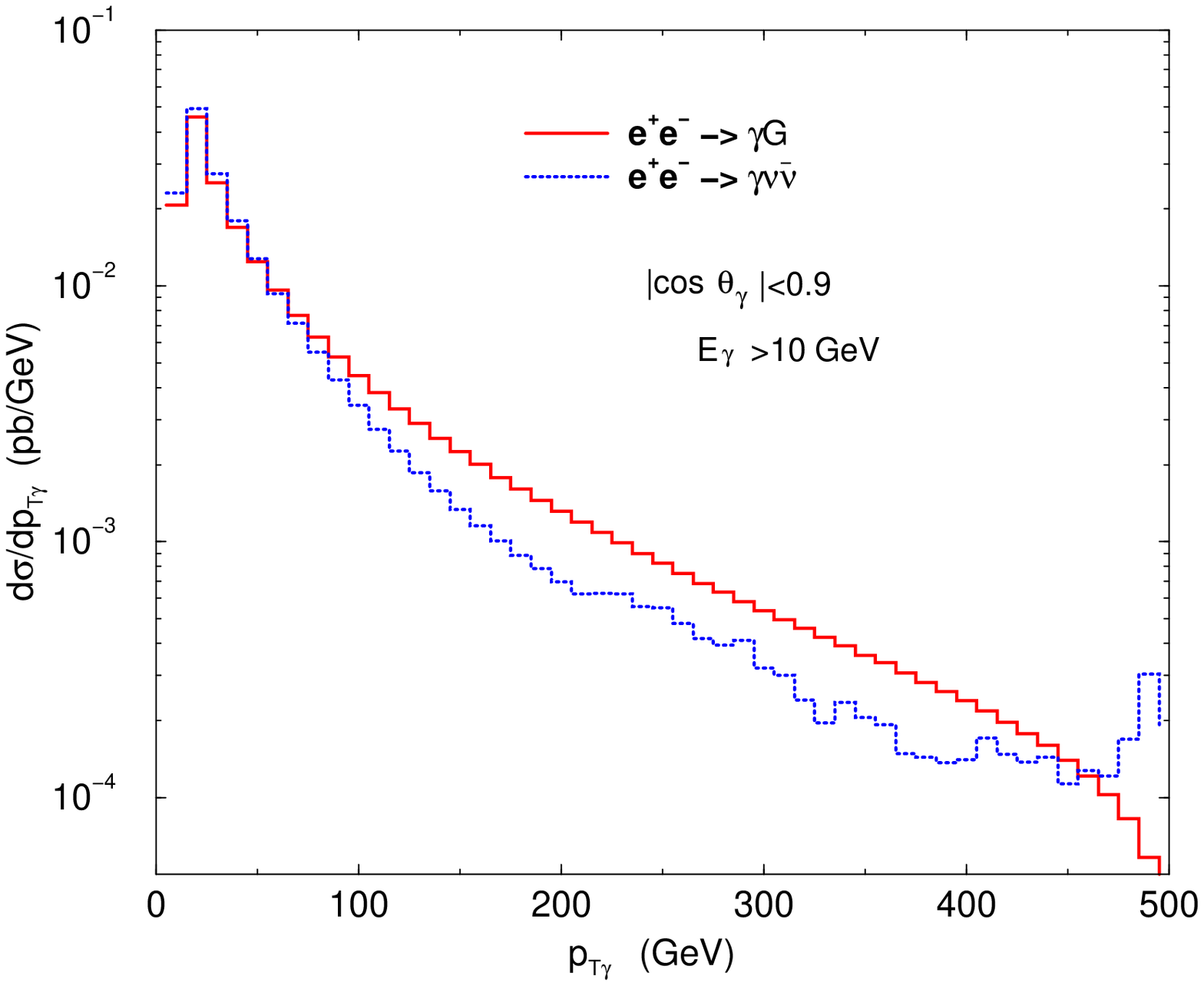}
\end{center}
\caption{(a) The differential distribution $d\sigma /d M_{\rm recoil}$ and
(b) $d\sigma /d P_{T_\gamma}$ 
for $e^+ e^- \to \gamma \nu_i \bar\nu_i$ $(i=e,\mu, \tau)$ 
and $e^+ e^- \to \gamma G$
for $n=2$ and $M_S=2.5$ TeV.  
We imposed cuts of $|\cos\theta_\gamma|<0.9$ and $E_\gamma >10$ GeV.
 }
\label{fig4}
\end{figure}

\end{document}